# Analysis of distracted pedestrians' waiting time: Head-Mounted Immersive Virtual Reality application

**Arash Kalatian[1], Anae Sobhani[2], Bilal Farooq[3]**
[1,3] Ryerson University
Toronto, Ontario
arash.kalatian@ryerson.ca; bilal.farooq@ryerson.ca
[2] Utrecht University
Utrecht, Netherlands
a.sobhani@uu.nl

***Abstract*** *–* This paper analyzes the distracted pedestrians' waiting time before crossing the road in three conditions: 1) not distracted, 2) distracted with a smartphone and 3) distracted with a smartphone in the presence of virtual flashing LED lights on the crosswalk as a safety measure. For the means of data collection, we adapted an in-house developed virtual immersive reality environment (VIRE). A total of 42 volunteers participated in the experiment. Participants' positions and head movements were recorded and used to calculate walking speeds, acceleration and deceleration rates, surrogate safety measures, time spent playing smartphone game, etc. After a descriptive analysis on the data, the effects of these variables on pedestrians' waiting time are analyzed by employing a cox proportional hazard model. Several factors were identified as having impact on waiting time. The results show that an increase in initial walk speed, percentage of time the head was oriented toward smartphone during crossing, bigger minimum missed gaps and unsafe crossings resulted in shorter waiting times. On the other hand, an increase in the percentage of time the head was oriented toward smartphone during waiting time, crossing time and maze solving time, means longer waiting times for participants.

## 1. Introduction

Pedestrian crossing behaviour is a topic of interest as it gives insights into traffic lights design, pedestrian safety, roadway layouts design and traffic flow optimization. This study aims at exploring one of the attributes of pedestrian street-crossing behaviour: waiting time. Research on pedestrian waiting time analysis have gained popularity as pedestrian-vehicle accidents result in a large proportion of total accident deaths. Pedestrian violations, i.e. J-waling, disobeying pedestrian lights and failure to yield to vehicles, have been a major source of traffic injury and fatality in recent years [1]. Despite implementation of various safety measures, widespread educational programs, etc., rate of pedestrian-related fatalities and injuries in Canada has increased in the last decade. According to the Canadian Motor Vehicle Traffic Collision Statistics, proportion of fatal accidents involving pedestrians to all fatal accidents has increased from 11.8% in 2005 to 15.2% in 2015 [2,3]. A reason for this increase can be traced back to the rise of smartphones and their applications in everyday life. Pedestrians are becoming more distracted in recent years, using their phones for talking, texting, surfing the web, looking for directions, or playing games [1].

In order for a pedestrian to cross an unsignalized intersection, individuals should wait for a gap that, based on each pedestrian's abilities, allows safe crossing of the street. An individual waiting to cross an unsignalized intersection is required to concentrate and evaluate whether each gap satisfies the spatial and temporal requirements of a safe cross. Pedestrian waiting time has a significant impact on unsafe crossings. Studies suggest a positive correlation between waiting time before initiating a cross and the violations caused by pedestrians [4]. Using mobile phones and the distraction caused by them negatively affects pedestrians' ability to cross and thus, increases the pedestrian-related accident rates [5,6]. By implementing safety solutions, waiting time can be affected for distracted pedestrian.

This paper investigates the effect of smartphone distraction on pedestrians' waiting time by adapting a head mounted immersive virtual reality environment. In conventional field experiments on pedestrian behaviours, it is often difficult to implement different scenarios as the results may be disastrous in terms of participants' safety. In addition, it may be expensive or unsafe to repeat an experiment with the exact same traffic conditions to capture the effects of an implemented safety measure. To overcome such problems, pictures, videos and photomontages can be used to assess participants perception. With the development of interactive computer-generated experiences, Virtual Reality (VR) experiments have gained popularity in various research fields. Using a head mounted VR display device, virtual scenes can be directly projected to the participants so that the they will be immersed in the simulated environment [7]. Scenarios used in experiments may be unsafe or expensive to apply on real roads, due to reasons such as dangerous implementations or lack of infrastructures. VR simulator allows running such scenarios, along with scenarios containing new technologies or services that participants have limited mental image of.

Data collection procedure for this study involves a virtual immersive reality environment (VIRE). To investigate distracted pedestrians' waiting time before crossing an unsignalized intersection, we asked participants to cross a simulated unsignalized intersection in VIRE in three different simulation tasks: Task 1. no distraction, Task 2. distraction by a maze on a smartphone, and Task 3. distraction by a maze game on a smartphone with flashing LED lights installed on the crosswalk as a safety encounter. After data collection and conducting a descriptive analysis of the data, a cox proportional hazard model is adopted to further explore the effects of different sociodemographic and traffic parameters on pedestrians' waiting times.

This paper is organized as follows: in Section 2, we review the previous works on the subject. In Section 3 data collection procedure is described. Section 4 discusses the model structure. Moreover, variables derived from data and a descriptive analysis on them are then explained in Section 5. Proportional Hazard model and its implementation on the data is then elaborated in Section 6. In the end, conclusions and future research plans are covered in Section 7.

## 2. Background

Despite the growing interest in pedestrian behaviour in the literature, pedestrian waiting time at intersections especially for unsignalized and mid-block crosswalks has yet to be widely discussed. For unsignalized intersections, Hamed [8] developed a cox proportional hazard model to identify factors affecting waiting time and number of unsuccessful attempts required before a safe crossing. Their Results suggest that having accident experience, car ownership, number of people on the crosswalk, age, gender, type of trip and vehicle gap time are important factors in determining wait time before crossing. However, waiting time at signalized intersections has drawn more attention in the literature, mainly by looking into calculation of signal timings based on pedestrians' waiting times. In 2003, Keegan and O'Mahony [9] studied the effects of different countdown timers on pedestrian waiting time at signalized intersections. In 2013, Li [10] developed a model for intended waiting time at signalized intersections taking into account bounded waiting times. Results showed that a large proportion of pedestrians cross the street immediately after they arrive at the intersection. In terms of the effects of waiting time on pedestrian violations, Brosseau *et al.* [4] analysed video data of signalized intersections and identified several factors that contribute to dangerous crossing, including maximum waiting time and clearance time.

Data in the aforementioned studies were collected either by questionnaires or by observing pedestrians for a short period of time. In addition, most studies on the subject consider waiting time at signalized intersections, mainly to explore the thresholds of pedestrian crossing signal timings. However, as it may be unsafe to track the pedestrian behavior while they are distracted using their phone, the distraction caused by smartphones has not been studied widely in previous literature.

Recent developments in virtual reality technology have made it possible to analyze different behaviours of pedestrians with minimal risks. Studies suggest that spatial knowledge developed in virtual environments resemble that of the actual physical environments [11,12]. Virtual Reality has been used in transportation studies in fields such as route choice or evacuation behaviour. However, these studies often lack the interaction element that can be implemented in Virtual Immersive Reality Environments (VIRE). Using VIRE, participants are immersed in an environment where vehicles respond to their actions. For instance, when a pedestrian walks in front of a car, the cars start to slow down, and if necessary, stop for the pedestrian to cross. Although experiments conducted in laboratory environments may lack the realism that is necessary for data collection, VIRE provides researchers with a safe environment that is close to real life situations. Considering the advantages of utilizing virtual reality environment for pedestrian-related experiments, data collection procedure was designed based on a virtual reality-based tool developed by LiTrans and introduced in [7].

## 3. Data Collection

In this study, an in-house developed tool, Virtual Immersive Reality Environment (VIRE) [7], is used to collect participants' data while they were waiting on the sidewalk to cross an unsignalized intersection. The simulated unsignalized intersection is part of an existing congested street in Montreal, Canada. After an initial familiarization with VIRE, participants were asked to engage in three waiting conditions mentioned earlier. To capture the repetition concern, each task was conducted in 10 random simulation scenarios. Each trial would finish when the participant successfully crosses the street, with maximum allowable time of 60 seconds [13]. Out of 42 participants, 17 were Females, the average age was 26, with 9 participants aged more than 30. Different variables were then generated based on participants collected data. As for the traffic simulation, car-following model and social force model were implemented for vehicular and pedestrian traffic, respectively. In Fig. 1, VIRE set-up and a couple of participants in the experiment can be seen.

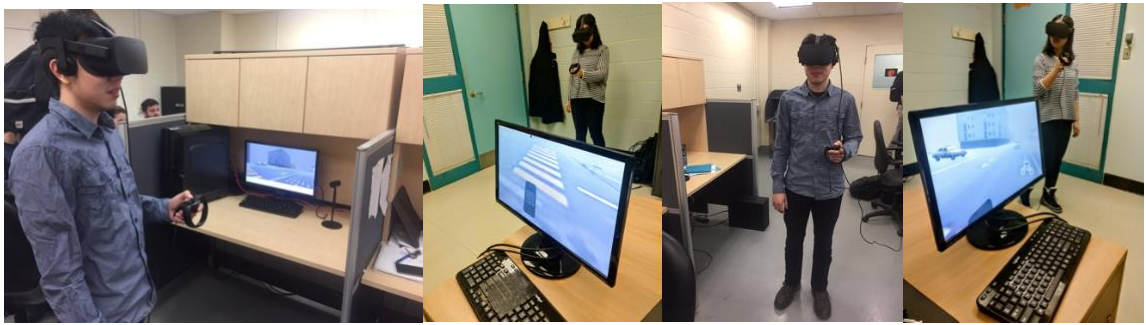

Fig. 1: VIRE setup and environment

## 4. Model Structure

Developed by Cox in 1972, Cox proportional-hazards model is essentially a regression model mainly used in medical research to identify the relationship between survival times of patients and predictor variables [14]. In this study, hazard function, denoted by ξ(*t*), is defined as the rate of failure to initiate a cross at time *t*. Hazard function is written as:

$$\xi(t) = \frac{\lim \Pr[t \leq T \leq t+\Delta t | T \geq t]}{\Delta t} \tag{1}$$

To analyze waiting time before a cross, cox proportional hazard model is used, as the most common method for analyzing individuals' survival. Considering the effects of covariates on the baseline hazard, hazard function is written as:

$$\xi(t|R) = \overline{\xi}(t) \ e^{\sum_{i=1}^{k} \chi_i R_i} \quad \text{and} \quad 0 \leq \xi(t|R) < \infty \tag{2}$$

In Eq. (2), R is the vector of covariates, $\chi$ is the vector of coefficients that need to be estimated, and $\bar{\xi}(t)$ is the baseline hazard [8]. Equation (2) gives the risk at time $t$ for pedestrian $i$, where the baseline hazard expresses hazard or risk for a pedestrian at all time points regardless of the covariates (R=0). To estimate model parameters, partial derivatives of the log-likelihood function is taken. As all the pedestrians finally cross the street at some point, our sample data is considered to be uncensored. To estimate parameters of the model, package "survival" in R is used in this study [15].

## 5. Data

Five types of variables were generated from the collected crossing behaviour data: 1) crossing variables: crossing duration (time from initiating the crossing until the crossing was completed), waiting time duration (the time from the start of the trial where the participant was standing on the sidewalk until the crossing was initiated), crossing speed, and initial walk speed, 2) distraction attributes: percentage of the time the head was facing the smartphone during wait time, percentage of the time the head was facing the smartphone while crossing, number of times the head was facing the smartphone over the total trial duration, maze solving time (the time it takes for a participant to solve one maze in the smartphone) , 3) safety measures: TTC (calculated for each second between vehicles and the pedestrian as the time until they would collide if their direction and speed remain unchanged), PET (the time difference between when the pedestrian departs the conflict point and the time vehicle arrives at that point), average and maximum acceleration and deceleration, the percentage of successful and time-out trials, 4) sociodemographic information: age, gender, number of years they owned a smartphone, having a driving license, and 5) gap variables: missing gaps before initiating crossing, and chosen gap for crossing. Table 1 presents the mean value of the attributes. As it can be seen in this Table, in the sample data collected, waiting time increases when participants are distracted. In the sample being studied, LED lights have caused females to wait slightly longer before crossing while distracted. However, Males waiting time has decreased a bit using LED lights. In all three scenarios, females have longer waiting times.

Table 1: Pedestrian behavioural attributes across three crossing conditions

| Variable | Task | General | Female | Male |
|---|---|---|---|---|
| Waiting time (s) | 1 | 18.0 | 20.9 | 16.0 |
|  | 2 | 21.0 | 23.9 | 19.4 |
|  | 3 | 21.0 | 24.3 | 19.2 |
| Crossing time (s) | 1 | 4.4 | 3.9 | 4.7 |
|  | 2 | 4.2 | 4.1 | 4.3 |
|  | 3 | 3.8 | 3.7 | 3.8 |
| Minimum gap missed (s) | 1 | 2.72 | 2.67 | 2.76 |
|  | 2 | 2.68 | 2.66 | 2.70 |
|  | 3 | 2.67 | 2.64 | 2.69 |
| % of time head was oriented toward smartphone during waiting time (s) | 2 | 72.9 | 68.0 | 76.2 |
|  | 3 | 74.7 | 69.8 | 78.0 |
| % of time head was oriented toward smartphone during crossing (s) | 2 | 73.5 | 73.1 | 73.7 |
|  | 3 | 69.6 | 71.5 | 68.3 |
| Initial walk speed (m/s) | 1 | 1.6 | 1.5 | 1.6 |
|  | 2 | 1.5 | 1.4 | 1.6 |
|  | 3 | 1.6 | 1.6 | 1.6 |
| Minimum PET (s) | 1 | 1.2 | 1.1 | 1.2 |
|  | 2 | 1.0 | 1.1 | 1.0 |
|  | 3 | 1.1 | 1.2 | 1.1 |
| Crossing success rate % | 1 | 76.9 | 7.6 | 81.2 |
|  | 2 | 61.4 | 64.7 | 59.2 |
|  | 3 | 68.8 | 65.9 | 70.8 |
| Maximum acceleration $m/s^2$ | 1 | 6.1 | 4.8 | 6.9 |
|  | 2 | 5.1 | 4.5 | 5.5 |
|  | 3 | 5.2 | 5.5 | 5.0 |
| Maze solving time (s) | 2 | 0.81 | 0.76 | 0.87 |
|  | 3 | 0.87 | 0.84 | 0.92 |

## 6. Model Results

To capture changes in waiting time pattern across the three conditions, and also estimate the effects of generated variables on waiting time, cox proportional hazard model is adopted. It should be noted that, in addition to estimating a variable's impact on waiting time, deviations of reasonable variables were evaluated and estimated through interaction variables, for example (Female) * (Maximum acceleration), (Female) * (Percentage of the time the head was oriented toward smartphone during waiting time), etc. Model specifications were calculated through a systematic process of removing statistically insignificant variables and combining variables when their effects were not significantly important. In safety analyses, survival analyse is the most common practice in modeling. In this study, we fit multivariate cox proportional hazard model to analyze the joint impact of factors on survival time. To better fit the model, dummy variables in different ranges were created for continues variables. Each variable is categorized in two or three groups, based on the distribution of the variable. Category thresholds were set in a way that each group contains at least 25 percent of participants. Crossings with a PET of less than 1.5 seconds were considered as dangerous or unsafe [16]. Table 2 provides estimated parameters of the significant covariates. The first column in this table presents variables that appeared to be significant. It should be noted that a positive coefficient sign means that the chance of starting a cross is higher, and thus, waiting time is shorter. The hazard ratio greater than one, indicates that as the value of the variable increases, the event's hazard increases and thus waiting time decreases.

As it can be seen in Table 2, the Task variable presents the differences of waiting time for Tasks 2 and 3 and Task 1. The positive value of this variable shows that waiting time in the tasks with smartphone involved is less than waiting time for non-distracted Task 1, meaning that smartphone usage has resulted in less waiting on the sidewalk, but the LED treatment has smoothened this negative impact. The second variable which has a significant effect on pedestrians' waiting time at an unsignalized intersection is gender with a negative impact for males. However, gender is not significant for separate tasks, implying that smartphone distraction does not change the behaviour of waiting time for any gender. Additionally, results show that pedestrians with higher crossing speeds (i.e. more rush to cross) waited less on the sidewalk. Analysing the distraction parameters values in Table 2 (i.e. head orientations toward phone while either waiting on the sidewalk or crossing a street) indicates that distracted pedestrians while waiting, spend more time on the sidewalk before initiating a cross, especially when there is no LED safety treatment implemented. On the other hand, distracted pedestrians, even while crossing the street, waited less at the sidewalk. Gap times that participants missed before crossing is reflected in minimum gap missed variable. In general, in all three waiting tasks, longer gap times have led to less waiting time. In task 1, more crossing duration means more waiting time, but this is not necessarily true for Tasks 2 and 3. In Tasks 2 and 3, in which participants are distracted with their phones, crossing duration may increase due to smartphone usage while crossing. Dangerous crossing, i.e. PET below 1.5 seconds, has a significant effect on pedestrians' waiting time for Task 1. Individuals with safer crossing tend to wait longer in task 1. However, this variable is removed in Task 2 and Task 3, which may be because of longer waiting times due to phone usage, instead of waiting for safe gaps. In the end, for each participant, as the time it takes to solve a maze increases, waiting time increases as well. This may be linked to the lack of concentration on road crossing on those participants who solve the maze faster.

Table 2- Results of multivariate proportional hazard model

| Variable | | Coefficient | Hazard Ratio | p-value |
|---|---|---|---|---|
| Task | 2 | 3.08 | 21.86 | 0.001 |
| | 3 | 2.93 | 18.82 | 0.001 |
| Female | General | -0.55 | 0.58 | 0.118 |
| Initial walk speed > 1.9 m/s | General | 0.33 | 1.38 | 0.121 |
| % of time head was oriented toward smartphone during waiting time >85% | General | -1.16 | 0.20 | 0.001 |
| | Task 2 | -0.96 | 0.38 | 0.012 |
| % of time head was oriented toward smartphone during crossing > 85% | General | 0.68 | -1.97 | 0.009 |
| Minimum gap missed > 3.1 s | General | 2.64 | 14.01 | 0.001 |
| | Task 1 | 1.42 | 4.17 | 0.001 |
| | Task 2 | 1.57 | 4.83 | 0.001 |
| | Task 3 | 1.16 | 3.21 | 0.028 |
| Crossing time > 5 s | Task 1 | -0.98 | 0.37 | 0.089 |
| PET < 1.5 s | Task 1 | 1.08 | 2.94 | 0.002 |
| Maze solving time > 1.1 s | Task 2 | -1.21 | 0.30 | 0.003 |
| | Task 3 | -2.27 | 0.10 | 0.001 |

## 7. Conclusions and Future Work

By developing an immersive virtual reality-based method for collecting data of pedestrian waiting time in street crossing paradigm, data can be collected in a safe and controlled environment. To observe the effects of various distraction parameters on pedestrian waiting time, cox proportional hazard model was adopted for the modelling purpose.

With regards to the model, results show higher minimum missed gap, dangerous PET (i.e. less than 1.5 seconds), crossing duration, initial crossing speed, percentage of time head orientation towards the phone during wait time and crossing, and gender.

Our study is not without limitations which can be addressed in future studies. Eye movements, brain activity and heart beats can be measured as indicators of participants' stress and distraction level. Other types of distraction can also be added to the experiment. In terms of dataset, larger datasets can be collected and analysed to remove the possible biases. Other safety measures can also be explored and compared to each other to better analyse the effect of different safety measures. Finally, waiting time for different types of intersections can also be studied.